\documentclass[aps,pra,a4paper,twocolumn,noshowpacs,superscriptaddress,floatfix]{revtex4}
\usepackage{amsfonts}
\usepackage{amsmath}
\usepackage{amssymb}
\usepackage{array}
\usepackage{graphicx}
\usepackage{epstopdf}
\usepackage{multirow}
\usepackage{booktabs}
\usepackage{makecell}
\usepackage{setspace}
\usepackage{booktabs}
\usepackage{threeparttable}
\usepackage[figuresright]{rotating}

\newcommand{\PreserveBackslash}[1]{\let\temp=\\#1\let\\=\temp}
\newcolumntype{C}[1]{>{\PreserveBackslash\centering}p{#1}}
\newcolumntype{R}[1]{>{\PreserveBackslash\raggedleft}p{#1}}
\newcolumntype{L}[1]{>{\PreserveBackslash\raggedright}p{#1}}

\begin{document}

\title{A Simple approach for precision calculation of Bethe logarithm}

\author{San-Jiang Yang}
\affiliation{College of Physics and Electronic Science, Hubei Normal University, Huangshi 435002, China}
\affiliation{State Key Laboratory of Magnetic Resonance and Atomic and Molecular Physics, Wuhan Institute of Physics and Mathematics, Innovation Academy for Precision Measurement Science and Technology, Chinese Academy of Sciences, Wuhan 430071, People’s Republic of China}

\author{Jing Chi}
\affiliation{State Key Laboratory of Magnetic Resonance and Atomic and Molecular Physics, Wuhan Institute of Physics and Mathematics, Innovation Academy for Precision Measurement Science and Technology, Chinese Academy of Sciences, Wuhan 430071, People’s Republic of China}
\affiliation{University of Chinese Academy of Sciences, Beijing 100049, People’s Republic of China}

\author{Wan-Ping Zhou \footnote{electronic mail: zhouwp@whu.edu.cn}}
\affiliation{School of Physics and Telecommunications, Huanggang Normal University, Huanggang 438000, China}

\author{Li-Yan Tang}
\affiliation{State Key Laboratory of Magnetic Resonance and Atomic and Molecular Physics, Wuhan Institute of Physics and Mathematics, Innovation Academy for Precision Measurement Science and Technology, Chinese Academy of Sciences, Wuhan 430071, People’s Republic of China}

\author{Zhen-Xiang Zhong}
\affiliation{Center for Theoretical Physics, School of Physics and Optoelectronic Engineering, Hainan University, Haikou 570228, China}

\author{Ting-Yun Shi \footnote{electronic mail: tyshi@wipm.ac.cn}}
\affiliation{State Key Laboratory of Magnetic Resonance and Atomic and Molecular Physics, Wuhan Institute of Physics and Mathematics, Innovation Academy for Precision Measurement Science and Technology, Chinese Academy of Sciences, Wuhan 430071, People’s Republic of China}

\author{Hao-Xue Qiao}
\affiliation{Department of Physics, Wuhan University, Wuhan 430072, People’s Republic of China}

\begin{abstract}
In this article we propose a simple approach for the precision calculation of Bethe logarithm. The leading contributions are obtained using specific operators, while the remaining terms are eliminated by adjusting the parameter $\lambda$. Through the use of dimensional regularization, singular divergences are algebraically canceled. Compared to the standard form of Bethe logarithm, our approach significantly reduces the complexity of constructing pseudostates in numerical evaluations. Using this approach we obtain a very highly precise result of Bethe logarithm for the ground state of the hydrogen, achieving 49 significant digits. And for multi-electron systems this approach appears simplicity and efficiency as well.
\end{abstract}

\keywords{Bethe logarithm, precision calculation, few-body systems}

\maketitle

\section{Introduction}
The interest in precision measurement and calculation on few-body atomic and molecular systems is greatly increasing due to the high precise determination for physical quantities, such as fine-structure constant $\alpha$ \cite{Schwartz1964, pachuckiFineStructureHeliumlike2010, ZhengLSFSS2017}, Rydberg constant \cite{JentschuraFCTTRS2008, Tan2014}, electron-proton mass ratio \cite{SayanProtonelectronMassRatio, KorobovHS2+, KorobovRSTH2+}, Zemach radii \cite{YerokhinHSLi2008, PuchalskiGSHSLi2013, QiPCHSZLi2020, SunMHSZR2023} and negatively charged pion mass \cite{HoriProposedMethodForLaserSpectroscopy, HoriLaserSpectroscopyofPionicHelium, BaiPrecisionSpectroscopyOfThePionicHelium4}. High precision quantities could provide calibration data for other researches, moreover, comparison between different approaches could accurately test quantum electrodynamics in atomic and molecular systems.

In spectral calculation Bethe logarithm correction is one quite remarkable term, which was firstly carried out by Hans Bethe in 1947 to explain the splitting between 2S and 2P energy levels of hydrogen \cite{LambFSofH, betheElectromagneticShiftEnergy1947}. Different from other correction operators, Bethe logarithm involving unperturbed Hamiltonian operator $H_0$ in logarithmic function,
\begin{equation}
\label{BLH}
\begin{aligned}
     \beta_{\rm H} = \frac{\left\langle \vec{p} (H_0-E_0) \ln [ 2(H_0-E_0) ] \vec{p} \right\rangle}{\left\langle \vec{p} (H_0-E_0) \vec{p}\right\rangle} \, ,
\end{aligned}
\end{equation}
which makes this term intractable in precise calculations. By applying the completeness relation Bethe logarithm appears more easily to deal with. However extremely slow convergence rate makes usual method hard to meet requirements, e.g., to obtain 4 decimal digits of $\beta_H$ the length of Slater-Laguerre basis set needs to reach around 27000 \cite{goldmanHighAccuracyAtomic2000}! Such a demand for calculation of Bethe logarithm in multi-electron systems is unrealistic. For the case of hydrogen-like systems, there is an impressive method based on $SO(4,2)$ providing an efficient approach to high accuracy numerical evaluation \cite{huffSimplifiedCalculation}. Especially for hydrogen ground state, Gavrila presented the analytical expression for Kramers-Heisenberg matrix element by using Schwinger integral representation of Green's function for the Coulomb field \cite{huffSimplifiedCalculation}. Additionally, in numerical evaluation the pseudostates could dramatically accelerate the convergence rate of Bethe logarithm, e.g., with exquisite construction of variational basis Goldman obtained 8 significant digits of $\beta_{\rm H}$ with only 20 basis functions \cite{goldmanIterationvariationalMethodIts1984}, and Goldman and Drake got 23 effective digits of ground state $\beta_{\rm H}$ under $E_{\rm max}$ arrived $10^{46}$ a.u. with total number of basis set only 377 \cite{goldmanHighAccuracyAtomic2000}. Inspired by Goldman and Drake's work, Tang et. al. constructed pseudostates using B-spline basis set, which works very well for Bethe logarithm calculation on highly-excited Rydberg states of hydrogen \cite{tangBethelogarithmCalculationUsing2013}.

For multi-electron systems Bethe logarithm was the main obstacle in precision calculation. An important approach was introduced by Schwartz in 1961 \cite{Schwartz1964}, which based on an integral form of Bethe logarithm,
\begin{equation}
\begin{aligned}
     \beta_{\rm He} = \frac{ \lim\limits_{K\rightarrow \infty} \left[ \langle\nabla^2\rangle K + 2\pi Z \psi_0^2(0)\ln K + \int_0^K k dk J(k) \right] }{ 2\pi Z \psi_0^2(0) }
\end{aligned}
\end{equation}
with $J(k) = -\langle \psi_1 | \nabla | \psi_0 \rangle$, and $\psi_1$ satisfies
\begin{equation}
\begin{aligned}
     (E_0-H_0-k) \psi_1 = \nabla \psi_0 \, .
\end{aligned}
\end{equation}
Schwartz divided integration interval into two parts and detailed analyzed the asymptotic behaviour of $\beta_{\rm He}$, and then provided 4 significant digits result for the ground state of helium. Remarkably this result remained as the best for over 30 years. In 1999 Drake and Goldman made the key breakthrough with successful construction of pseudostates for helium-like systems \cite{drakeBetheLogarithmsPs2000}. By employing pseudostates they improved the accuracy of Bethe logarithm computations for helium-like systems to around 10 significant digits. Lately this approach was successfully generalized to lithium by Yan and Drake \cite{yanBetheLogarithmQED2003}. In the same year, Korobov and Korobov extended Schwartz's method and calculated Bethe logarithm of helium \cite{korobovBetheLogarithmStates1999}. In their treatment the integration interval of virtual photon energy is divided into three parts. The asymptotic coefficients in high energy part are obtained by a fitting method. Lately Korobov, Korobov and Zhong extand this approach to $H_2^+$ and $HD^+$ calculation \cite{korobovCalculationNonrelativisticBethe2012, korobovBetheLogarithmHD2012}. And Korobov, Hilico and Karr use this approach in calculation of relativistic Bethe logarithm in two-center problem \cite{korobovCalculationRelativisticBethe2013}. Meanwhile the asymptotic analysis has also been used by Pachucki to calculate two-loop Bethe logarithm for hydrogen \cite{pachuckiLogarithmicTwoloopCorrections2001}, by Pachucki and Komasa in calculation of Bethe logarithm for lithium \cite{pachuckiBetheLogarithmLithium2003a}, by Pachucki and Yerokhin in relativistic correction to Bethe logarithm calculation for helium \cite{pachuckiFineStructureHeliumlike2010}, and by Yerokhin, Patkóš and Pachucki in spin-independent relativistic correction to Bethe logarithm \cite{YerokhinRBL2018}.

In this article we focus on nonrelativistic atomic Bethe logarithm calculation. Through nonrelativistic quantum electrodynamics (NRQED) approach \cite{CASWELL1986437, pachuckiSimpleDerivationHelium1998, pachuckiCorrectionsSingletStates2006} we present a rapid convergence form for Bethe logarithm calculation. The leading asymptotic coefficients are obtained by the expectation values of some operators. And with the help of dimensional regularization the divergent parts are canceled algebraically. For the remaining asymptotic coefficients we found it is not necessary in our calculation. By adjusting parameter $\lambda$ we could obtain precision results efficiently with one diagonalization. This article arranged as follows. In Sec. \ref{HYDROGEN} we first derive the rapid convergence form of Bethe logarithm (rcfBL) for hydrogen, regularizing operators in $d=3-2\epsilon$ dimensions and canceling divergent terms by combing high-energy parts. We then analyze the convergence pattern of rcfBL, and evaluate Bethe logarithm for the ground state of hydrogen with varying $\lambda$. Comparison between rcfBL and standard form of Bethe logarithm (sfBL) is available. In Sec. \ref{MULTI-ELECTRON ATOMIC SYSTEMS} we demonstrate rcfBL for multi-electron systems, providing detailed expressions of helium and corresponding numerical results of necessary operators for $1\;{}^1S$, $2\;{}^1S$ and $2\;{}^3S$ states. With these values we calculate Bethe logarithm for these states of helium to display the simplicity and efficiency of rcfBL. Finally a brief summary is given in Sec. \ref{Summary}.

\section{HYDROGEN}
\label{HYDROGEN}
Bethe logarithm could be seen as the contribution of low-energy virtual photons, $|\vec{k}|<m\alpha$. In this energy scale, with dipole approximation and using nonrelativistic Hamiltonian $H_0$, the energy shift of hydrogen could be written as, $e^2=4\pi\alpha$,
\begin{equation}
\begin{aligned}
\Delta E =& \frac{e^2}{m^2} \int^{\varepsilon m \alpha} \frac{d^3k}{(2\pi)^3 2k} \left( \delta^{ij}-\frac{k^ik^j}{k^2} \right) \times \\
&\left\langle p^i \left[ \frac{1}{E_0-H_0-k} + \frac{1}{k} \right] p^j \right\rangle \, ,
\end{aligned}
\end{equation}
where $e$ and $m$ are the charge and mass of electron respectively. $\alpha$ is the fine-structure constant, $\varepsilon$ represents the cut-off parameter less than 1 and $E_0$ is the energy level corresponding to initial state. After angular integration one obtain
\begin{equation}
\label{DeltaE}
\begin{aligned}
\Delta E = -\frac{e^2}{6\pi^2m^2} \left\langle \vec{p}(E_0-H_0)\ln \left( 1+\frac{\varepsilon m \alpha}{H_0-E_0} \right) \vec{p} \right\rangle \, .
\end{aligned}
\end{equation}
Noticing that in lower exited states of system $H_0$ is of order $\alpha^2$, which is smaller than $\epsilon m\alpha$ by one $\alpha$, in $m\alpha^5$, after dropping the divergent term proportional to $\ln(2\varepsilon / \alpha)$, one could get Bethe logarithm from
\begin{equation}
\begin{aligned}
\Delta E =& -\frac{2}{3\pi} \left\langle 2\pi \delta ^3(\vec{r}) \right\rangle \beta_{\rm H} \, .
\end{aligned}
\end{equation}

Compared to Eq. (\ref{BLH}), Eq. (\ref{DeltaE}) is less sensitive to high-energy intermediate states. In numerical evaluation we might start with Eq. (\ref{DeltaE}) to reduce the task of constructing intermediate states. More details, by performing an asymptotic expansion of Eq. (\ref{DeltaE}), we could present Bethe logarithm as $\mathcal{N}/\mathcal{D}$,
\begin{equation}
\label{N}
\begin{aligned}
\mathcal{N} =& \left\langle \vec{p}(E_0-H_0)\ln\mathrm{(}1+\frac{\lambda}{H_0-E_0})\vec{p} \right\rangle  \\
&+\left\langle \vec{p}(H_0-E_0)\vec{p} \right\rangle \ln (2\lambda)  \\
&+\frac{1}{\lambda}\left\langle \vec{p}(E_0-H_0)^2\vec{p} \right\rangle +\frac{1}{2\lambda ^2}\left\langle \vec{p}(E_0-H_0)^3\vec{p} \right\rangle +\cdots
\end{aligned}
\end{equation}
and
\begin{equation}
\label{D}
\begin{aligned}
\mathcal{D} = \left\langle \vec{p}(H_0-E_0)\vec{p} \right\rangle = 2\pi \left\langle \delta ^3(\vec{r}) \right\rangle  \, ,
\end{aligned}
\end{equation}
where $\lambda$ is a free parameter with no upper bound. Sufficient large $\lambda$ allows us to drop the third line of  $\mathcal{N}$, but this actually goes back to sfBL of hydrogen. In order to reduce the value of $\lambda$,  that is, to increase the convergence rate of $\mathcal{N}$, we deal with $\langle \vec{p}(E_0-H_0)^2\vec{p} \rangle$, denoted by $C_{\lambda}$, in the third line as follows. With the help of commutation relations $C_{\lambda}$ could be rewritten as
\begin{equation}
\begin{aligned}
\left\langle \vec{p}(E_0-H_0)^2\vec{p} \right\rangle = -\left\langle [\vec{p}, H_0]^2 \right\rangle \, .
\end{aligned}
\end{equation}
This term is divergent for S-states at low $r$, which should be regularized and be completed by high-energy part. Here we work in $d=3-2\epsilon$ dimensions (details see Ref. \cite{pachuckiCorrectionsSingletStates2006}). After recombining the commutators and utilizing the Schr\"{o}dinger equation for operator equivalence substitutions we separate singularity as
\begin{equation}
\label{C}
\begin{aligned}
C_{\lambda} =& \left\langle\vec{p}\frac{1}{r^2}\vec{p}\right\rangle 
-2E_0 \left\langle\frac{1}{r^2}\right\rangle 
-2 \left\langle\frac{1}{r^3}\right\rangle \\
&-2\pi \left\langle\delta ^d\left( r \right)\right\rangle  \left( \frac{1}{\epsilon}+2 \right) 
\, ,
\end{aligned}
\end{equation}
where $\langle 1/r^3 \rangle$ should be understood as \cite{ArakiQuantumElectrodynamicalCorrections, SucherEnergyLevels}
\begin{equation}
\label{Pr3}
\begin{aligned}
\left\langle\phi\left|\frac{1}{r^3}\right| \psi\right\rangle =& \lim _{a \rightarrow 0} \int d^3 r \phi^*(\vec{r}) \psi(\vec{r}) \times \\
&\left[\frac{1}{r^3} \Theta(r-a)+4 \pi \delta^3(r)(\gamma_E+\ln a)\right]
\, ,
\end{aligned}
\end{equation}
where $\gamma_E$ is Euler's constants. The $\epsilon$-dependence term will be canceled by combining with forward-scattering exchange amplitudes \cite{pachuckiLogarithmicTwoloopCorrections2001}. Two photon exchange is
\begin{equation}
\begin{aligned}
P_2(\omega) =& -128\pi ^2\psi_0^2(0) \int{\frac{d^dp}{\left( 2\pi \right) ^d}\frac{\vec{p}}{p^4}\frac{1}{p^2+2\omega}}\frac{\vec{p}}{p^4} \\
=& -4\pi \frac{\sqrt{2\omega}}{\omega ^3}\psi_0^2(0) 
\, ,
\end{aligned}
\end{equation}
and three photon exchange is $P_{3A}+P_{3B}$,
\begin{equation}
\begin{aligned}
P_{3A}(\omega) =& -1024\pi ^3\psi_0^2(0) \int{\frac{d^dp_1}{\left( 2\pi \right) ^d} \int{\frac{d^dp_2}{\left( 2\pi \right) ^d}}} \times
\\
& \frac{\vec{p}_1}{p_{1}^{4}}\frac{1}{p_{1}^{2}+2\omega} \frac{1}{\left( \vec{p}_1-\vec{p}_2 \right) ^2}\frac{1}{p_{2}^{2}+2\omega}\frac{\vec{p}_2}{p_{2}^{4}}
\\
=& 4\pi \frac{-1-\ln 2+\ln \omega}{\omega ^3}\psi_0^2(0) -\frac{1}{\omega ^3}\frac{2\pi}{\epsilon}\psi_0^2(0) 
\, ,
\end{aligned}
\end{equation}
and
\begin{equation}
\begin{aligned}
P_{3B}(\omega) =& -2048\pi ^3\psi_0^2(0) \times \\
& \int{\frac{d^dp_1}{\left( 2\pi \right) ^d} \int{\frac{d^dp_2}{\left( 2\pi \right) ^d}}} \frac{1}{p_{1}^{4}}\frac{1}{\left( \vec{p}_1-\vec{p}_2 \right) ^2}\frac{1}{p_{2}^{2}+2\omega}\frac{1}{p_{2}^{4}} \\
=& \frac{8\pi}{\omega ^3}\psi_0^2(0) 
\, .
\end{aligned}
\end{equation}
Through integration over transverse photon and dropping the coefficient $-2/3\pi$, the high-energy party reads
\begin{equation}
\label{S}
\begin{aligned}
\left( \frac{2\sqrt{2\lambda}-2+\ln 2-\ln \lambda}{\lambda} \right) \langle 4\pi \delta ^d\left( \vec{r} \right) \rangle +\frac{1}{\lambda}\frac{2\pi}{\epsilon}\langle \delta ^d\left( \vec{r} \right) \rangle 
\, .
\end{aligned}
\end{equation}
Summing up Eq. (\ref{N}), (\ref{C}) and (\ref{S}), the numerator of hydrogen Bethe logarithm could be presents as
\begin{equation}
\begin{aligned}
\mathcal{N} =& \left\langle \vec{p}(E_0-H_0)\ln\mathrm{(}1+\frac{\lambda}{H_0-E_0})\vec{p} \right\rangle  \\
&+\left\langle \vec{p}(H_0-E_0)\vec{p} \right\rangle \ln (2\lambda)  \\
&+ \frac{2\sqrt{2}}{\sqrt{\lambda}}\langle 4\pi \delta ^3\left( \vec{r} \right) \rangle + \frac{c_{\lambda}}{\lambda} + O(\frac{1}{\lambda^{3/2}})\, ,
\end{aligned}
\end{equation}
with
\begin{equation}
\begin{aligned}
c_{\lambda} =& \left\langle\vec{p}\frac{1}{r^2}\vec{p}\right\rangle 
-2E_0 \left\langle\frac{1}{r^2}\right\rangle 
-2 \left\langle\frac{1}{r^3}\right\rangle \\
&+ (\ln 2-3-\ln \lambda) \langle 4\pi \delta ^3\left( \vec{r} \right) \rangle
\, .
\end{aligned}
\end{equation}

To analyze the convergence pattern of rcfBL we calculate Bethe logarithm for the ground state of hydrogen. With $c_{\lambda} = 4(3\ln 2 -2 -\ln \lambda)$, we list our results in Table \ref{resBLH}. The pseudostates are generated by Slater basis functions,
\begin{equation}
\begin{aligned}
\phi_{ik}(\vec{r}) = r^k e^{-\alpha_i r} Y_{10}(\hat{r})  \, ,
\end{aligned}
\end{equation}
where $Y$ is the spherical harmonic function and $k\le 1$. The nonlinear parameters are generated according to the following formula, similar to the universal basis set \cite{malliUniversalGaussianBasis},
\begin{equation}
\begin{aligned}
\alpha_i = e^{-0.38 + (i-2)*0.36}  \, .
\end{aligned}
\end{equation}
To avoid numerical degeneracy problem we work in 256-digits decimal precision in our calculations \cite{Bailey}. The first column of this table displays the length of basis set, and the last column denotes the maximum energy of pseudostates. The second and the third column contain our results of Bethe logarithm for the ground state of hydrogen calculated by rcfBL with $\lambda=1\times 10^{10}$ and  $\lambda=1\times 10^{12}$ respectively. These results demonstrate that rcfBL quite fast reaches its upper limit of accuracy $\lambda^{-3/2}$ when $E_{\rm max}$ around $\lambda$. We also evaluate sfBL under the same pseudostates, and tabulated results in the fourth column. Compared to sfBL, rcfBL possesses a much more rapid convergence rate in numerical evaluation. Table \ref{resBLH2} contains our results obtained by rcfBL with varying $\lambda \sim E_{\rm max}/100$. In these calculations we set $k \le 5$. From this table we could found that with $\lambda$ grows, the number of significant digits of results increases dramatically, reaching up to 49 significant digits. Apparently it requires considerably more computational effort to reach similar levels of precision by using sfBL.

\section{MULTI-ELECTRON ATOMIC SYSTEMS}
\label{MULTI-ELECTRON ATOMIC SYSTEMS}
For the case of N-electron atomic systems, the $\vec{p}$ in Eq. (\ref{N}) and (\ref{D}) should be understood as $\vec{p} = \sum_i \vec{p}_i$. The leading term of asymptotic expansion becomes
\begin{equation}
\label{pH2p}
\begin{aligned}
\left\langle \vec{p} (E_0 - H_0)^2 \vec{p}\right\rangle = - \sum_{ij} \left\langle \left[ \vec{p}_i , \frac{Z}{r_i} \right] \left[ \vec{p}_j , \frac{Z}{r_j} \right] \right\rangle \, ,
\end{aligned}
\end{equation}
where $Z$ is the nuclear charge. The regularization of this term only involves coulomb interaction between electron and nucleus, and the separation of singularity is similar to the case of hydrogen. The corresponding high-energy part is obtained from hydrogen-like ions,
\begin{equation}
\begin{aligned}
Z^2 P_2(\omega) + Z^3 [P_{3A}(\omega) + P_{3B}(\omega)] \, ,
\end{aligned}
\end{equation}
which cancels out the divergence of Eq. (\ref{pH2p}). For the case of helium, we could express $\beta_{\rm He}$ as $\mathcal{N}_{\rm He}/\mathcal{D}_{\rm He}$,
\begin{equation}
\label{Nhe}
\begin{aligned}
\mathcal{N}_{\rm He} =& \left\langle \vec{p}(E_0-H_0)\ln\mathrm{(}1+\frac{\lambda}{H_0-E_0})\vec{p} \right\rangle  \\
&+\left\langle \vec{p}(H_0-E_0)\vec{p} \right\rangle \ln (2\lambda)  \\
&+ \frac{2\sqrt{2}Z^2}{\sqrt{\lambda}}4\pi \langle \delta ^3\left( \vec{r}_1 \right) + \delta ^3\left( \vec{r}_2 \right) \rangle + \frac{d_{\lambda}}{\lambda} + O(\frac{1}{\lambda^{3/2}})\, ,
\end{aligned}
\end{equation}
with
\begin{equation}
\label{dlambda}
\begin{aligned}
d_{\lambda} =& 4\pi \langle \delta ^3\left( \vec{r}_1 \right) +\delta ^3\left( \vec{r}_2 \right) \rangle Z^3 \left( -3+\ln 2-\ln \lambda \right) 
\\
&+4Z^2\left\langle \frac{1}{r_{1}^{2}r_{12}} \right\rangle
-4Z^2E_0\left\langle \frac{1}{r_{1}^{2}} \right\rangle
+2Z^2\left\langle p_{2}^{2}\frac{1}{r_{1}^{2}} \right\rangle  \\
&+2Z^2\left\langle p_1\frac{1}{r_{1}^{2}}p_1 \right\rangle
-4Z^3\left\langle \frac{1}{r_{1}^{2}r_2} \right\rangle
-4Z^3\left\langle \frac{1}{r_{1}^{3}} \right\rangle  \\
&-4Z^2 \left\langle \frac{1}{r_1 r_2} p_1 p_2 \right\rangle
+4Z^2 \left\langle  p_2 \frac{1}{r_1 r_2} p_1 \right\rangle 
\, .
\end{aligned}
\end{equation}
In our calculation the first term of $\mathcal{N}_{\rm He}$ is transformed into ``acceleration gauge'' using the commutation relation
\begin{equation}
\label{Acceleration}
\begin{aligned}
\left\langle \psi_0 | [\vec{p},E_0-H_0] |\psi_m \right\rangle = iZ \left\langle \psi_0 \left|  \frac{\vec{r}}{r^3} \right| \psi_m \right\rangle  \, .
\end{aligned}
\end{equation}
The definition of $\langle 1/r^3 \rangle$ in $d_{\lambda}$ is same to Eq. (\ref{Pr3}), and the denominator is
\begin{equation}
\label{Dhe}
\begin{aligned}
\mathcal{D}_{\rm He} = \left\langle \vec{p}(H_0-E_0)\vec{p} \right\rangle = 2\pi Z \left\langle \delta ^3(\vec{r}_1) + \delta ^3(\vec{r}_2)\right\rangle  \, .
\end{aligned}
\end{equation}

We turn to evaluate Bethe logarithm of helium based on Eq. (\ref{Nhe}), (\ref{dlambda}), (\ref{Acceleration}) and (\ref{Dhe}). In order to get sufficient accurate initial state we expand trial wavefunction in the following form
\begin{equation}
\begin{aligned}
\psi_{LM}(\vec{r}_1, \vec{r}_2) &= \\
\sum_{i=1}^{N} v_i &e^{-\alpha_i r_1 -\beta_i r_2 -\gamma_i r_{12}} \Lambda_{l_1 l_2}^{LM}(\hat{r}_1, \hat{r}_2) + (\vec{r}_1 \leftrightarrow \vec{r}_2) \, .
\end{aligned}
\end{equation}
Where $r_i$ is the distance between electron and nucleus, and $r_{12}$ is the distance between two electrons. $\Lambda_{l_1 l_2}^{LM}$ denotes the vector coupled product of  angular momenta $l_1$ and $l_2$ for the two electrons to form the $L(L+1)$ and $M$ eigenstates. The total number of basis set is restricted by $N$, and $\alpha_i$s, $\beta_i$s and $\gamma_i$s are the nonlinear parameters which could be generated in a quasirandom manner \cite{frolovUniversalVariationalExpansion1995, KorobovCoulombTthreeNodyBoundStateProblem},
\begin{equation}
\begin{aligned}
\alpha _i= \left\lfloor \frac{1}{2}i(i+1)\sqrt{2} \right\rfloor \left( A_2-A_1 \right) +A_1 \, ,
\\
\beta _i=  \left\lfloor \frac{1}{2}i(i+1)\sqrt{3} \right\rfloor \left( B_2-B_1 \right) +B_1 \, ,
\\
\gamma _i= \left\lfloor \frac{1}{2}i(i+1)\sqrt{5} \right\rfloor \left( C_2-C_1 \right) +C_1 \, .
\end{aligned}
\end{equation}
Here $\lfloor x \rfloor$ designates the fractional part of $x$. The interval $[A_1, A_2]$ is the range for generating nonlinear parameters $\alpha_i$s, and similarly for $[B_1, B_2]$ and $[C_1, C_2]$. To improve computational efficiency the quasirandom-generated $\alpha_i$s, $\beta_i$s and $\gamma_i$s are further optimized in our calculation. The calculation of hamiltonian $H_0$ and overlap $O$ matrices are based on the basic formula
\begin{equation}
\begin{aligned}
\int d^3 r_1 \int d^3 r_2 \frac{e^{-\alpha r_1 -\beta r_2 -\gamma r_{12}}}{r_1 r_2 r_{12}} = \frac{16 \pi^2}{(\alpha + \beta)(\alpha + \gamma)(\beta + \gamma)} \, .
\end{aligned}
\end{equation}
The expressions with additional polynomial of $r$ could be obtained by differentiating with respect to the corresponding nonlinear parameter. The integrals involving negative power of $r$ are more complicated, which could be obtained from the following formulas,
\begin{equation}
\begin{aligned}
\int{d^3}r_1\int{d^3}r_2\frac{e^{-\alpha r_1-\beta r_2-\gamma r_{12}}}{r_{1}^{2}r_2r_{12}}=\frac{16\pi ^2}{\beta ^2-\gamma ^2}\ln \frac{\alpha +\beta}{\alpha +\gamma} \, ,
\end{aligned}
\end{equation}
and
\begin{equation}
\begin{aligned}
\int{d^3}r_1\int{d^3}r_2\frac{e^{-\alpha r_1-\beta r_2-\gamma r_{12}}}{r_{1}^{2}r_{2}^{2}r_{12}}=\frac{8\pi ^2}{\gamma}\left[ \frac{1}{2}\ln ^2\left( \frac{\alpha +\gamma}{\beta +\gamma} \right) \right. \\
\left. +{\rm dilog} \left( \frac{\alpha +\beta}{\alpha +\gamma} \right) +{\rm dilog} \left( \frac{\alpha +\beta}{\beta +\gamma} \right) +\frac{\pi ^2}{6} \right]  \, .
\end{aligned}
\end{equation}
More details could be found in Ref. \cite{harrisSingularNonsingularThreebody2004}. To avoid the strict requirement of $\delta^3(\vec{r}_i)$ operators, that is, the expectation value of this operator only depends on the information of wavefunctions at origin, we use the following transformation in our calculations \cite{HillerNewTechniques, Drachman1981},
\begin{equation}
\begin{aligned}
4\pi \langle \delta ^3\left( \vec{r}_1 \right) \rangle &= 4\left\langle \frac{E}{r_1}+\frac{Z}{r_{1}^{2}}+\frac{Z}{r_1r_2}-\frac{1}{r_1r_{12}} \right\rangle \\
&-2\left\langle p_{2}^{2}\frac{1}{r_1}+p_1\frac{1}{r_1}p_1 \right\rangle  \, .
\end{aligned}
\end{equation}

We evaluate eigenvalues of $H_0$ with $N=600,1200,1800,2400$ respectively. The results for $1\;{}^1S$, $2\;{}^1S$ and $2\;{}^3S$ states of helium are summarized in Table \ref{operators}. The expectation values of necessary operators in our calculation evaluated within the obtained wavefunctions are also included in this table. These results agree very well with other precision calculations \cite{Drake2005SpringerHO, pachuckiCorrectionsSingletStates2006, patkosHigherorderRecoilCorrections2017}. With these values we calculated Bethe logarithm for $1\;{}^1S$, $2\;{}^1S$ and $2\;{}^3S$ states of helium based on Eq. (\ref{Nhe}), (\ref{dlambda}), (\ref{Acceleration}) and (\ref{Dhe}).

The convergence study of Bethe logarithm for the ground state of helium is listed in Table \ref{resBLHe}. The initial state wavefunction is the same to the wavefunction used in the previous calculation on expectation values of operators with $N=2400$. The pseudostates contains two type basis sets. The first type $\{ \phi^A \}$ is constructed with triple-fold nonlinear parameters to precisely describe the electrons coalescences. The nonlinear parameters $\alpha$s, $\beta$s and $\gamma$s of the first fold are generated in $[0.1, 5]$, $[0.1, 5]$ and $[0, 0.05]$ respectively. The ranges to generated parameters of the second and the third fold are 3 times and 7 times that of the first fold. Similar to the calculation of initial states further optimization is necessary. The number of basis functions of each fold are same, here we denote it by $N_A$. The second type basis set $\{ \phi^B \}$ is determined by the first fold of $\{ \phi^A \}$. Specifically $\{ \phi^B \}$ consists of $k$-folds. For the $i$th fold the nonlinear parameters $\alpha^{B_i}$s and $\gamma^{B_i}$s are same to the first fold of $\{ \phi^A \}$, the $\beta^{B_i}$s are determined by $\beta^{B_i} = \beta^{A_1} \times 5^i$. In our calculation $k=1,2,\cdots, 6$, and we set the number of basis functions of each fold $N_B = N_A/2$. In our calculation we choose $\lambda=1\times 10^6$ and $\lambda=1\times 10^8$ to evaluate Bethe logarithm for the ground state of helium, with $N_A=100,200,300,400,500$. These results are listed in the second and the third column in Table \ref{resBLHe} respectively. Comparing the results from these two columns, we estimate that for the case of helium the influence of unknown higher-order terms would reduce one significant digits to $\lambda^{-3/2}$. From the results listed in the third column we conclude that the numerical result of our calculation is $\beta_{\rm He}(1\;{}^1S) = 4.370 160 222 9(1)$ a.u., which agrees very well with other precision results. And compared to the result obtained in sfBL under the same pseudostates, the result calculated in rcfBL is more accurate by 4 significant digits. The same pseudostates are also used to calculate Bethe logarithm for the exited state, $2\;{}^1S$, of helium. The similar convergence pattern are obtained, see Table \ref{resBLHe2}. This table also contains the numerical results for the triplet state, $2\;{}^3S$, of helium. The pseudostates are constructed in the same strategy. The results of these two states also converge to 11 significant digits. Although rcfBL provides a rapid convergence rate for Bethe logarithm calculation in N-electron atomic systems, the competitive relationship between the energy range and the description of electron correlations during the construction of intermediate states still poses a significant challenge for achieving very high precision results. Obtaining higher-order coefficients analytically would be beneficial. Nevertheless the accuracy of these results already satisfies the current requirements.

\section{Summary}
\label{Summary}
From Bethe formulated the Bethe logarithm correction up to the present, the calculation on this term has remained an important topic in precision calculation. Utilizing information from asymptotic coefficients could effectively reduce the complexity of constructing intermediate states. In this work we demonstrates that, by adjusting parameter $\lambda$, only the leading terms needed to be handled analytically, which could significantly reduce the complexity of numerical calculations. Using this approach we achieve a very highly precise result of Bethe logarithm for the ground state of hydrogen. For the case of multi-electron atomic systems this approach retains its simplicity and efficiency, with the singular divergences being canceled algebraically. This approach provides an efficient way to Bethe logarithm calculation of lithium or beryllium, and the extension to calculation of molecular systems Bethe logarithm and of relativistic Bethe logarithm appears very promising.

\section{Acknowledgments}
The authors are grateful to Z.-C. Yan and L.-M. Wang for meaningful discussions on Bethe logarithm correction, and to Yan for many valuable suggestions. S.-J. Yang thanks X.-Q. Qi for helpful discussion on details of numerical calculation. This work is supported by the National Natural Science Foundation of China No. 12304271, No. 12074295, No. 12393821 and No. 12174402. L.-Y. Tang is also supported by the Chinese Academy of Sciences Project for Young Scientists in Basic Research under Grant No. YSBR-055. The numerical calculations in this article have been done on the APM-Theoretical Computing Cluster (AMP-TCC).

\makegapedcells
\setcellgapes{4pt}
\begin{table*}[!htbp]
\caption{Convergence study of Bethe logarithm for the ground state of hydrogen, with a shift factor $\ln 2$. Values within square brackets represent the exponential part with base 10. Units are a.u.}
\label{resBLH}
\begin{tabular}{L{1.5cm}L{4cm}L{4.2cm}C{3.5cm}L{1.5cm}}
\hline
\hline
N&\qquad $\lambda=1[10]$  &\qquad $\lambda=1[12]$  &standard form of BL   &\quad $E_{\rm max}$\\
\hline
50 &2.290 88 &2.290 84  &2.290 83 & 2.9[8] \\
60 &2.290 980 &2.290 96  &2.290 95 & 1.0[10] \\
70 &2.290 981 375 205 557 &2.290 980  &2.290 97 & 3.9[11] \\
80 &2.290 981 375 205 558 &2.290 981 375 201  &2.290 980 & 1.4[13] \\
90 &2.290 981 375 205 558 &2.290 981 375 205 552 293  &2.290 980 & 5.2[14] \\
100 &2.290 981 375 205 558 &2.290 981 375 205 552 293  &2.290 980 & 1.9[16] \\
ref. \cite{goldmanHighAccuracyAtomic2000} & \multicolumn{4}{l}{2.290 981 375 205 552 301 342 51} \\
ref. \cite{huffSimplifiedCalculation} & \multicolumn{4}{l}{2.290 981 375 205 552 301 342 544 968 6} \\
\hline
\hline
\end{tabular}
\end{table*}

\makegapedcells
\setcellgapes{4pt}
\begin{table*}[!htbp]
\caption{Convergence study of Bethe logarithm for the ground state of hydrogen in varying $\lambda$, with a shift factor $\ln 2$. Values within square brackets represent the exponential part with base 10. Units are a.u.}
\label{resBLH2}
\begin{tabular}{L{2.5cm}L{12cm}L{1.5cm}}
\hline
\hline
(N,\, $\lambda$)& rapid convergence form of BL &\quad $E_{\rm max}$\\
\hline
(120, 1[5] ) & 2.290 981 5                                                             & 6.6[7]     \\
(180, 1[8] ) & 2.290 981 375 21                                                        & 8.8[10]    \\
(240, 1[12]) & 2.290 981 375 205 552 307                                               & 1.1[14]    \\
(300, 1[15]) & 2.290 981 375 205 552 301 342 7                                         & 1.5[17]    \\
(360, 1[18]) & 2.290 981 375 205 552 301 342 544 97                                    & 2.1[20]    \\
(420, 1[21]) & 2.290 981 375 205 552 301 342 544 968 555 8                             & 2.8[23]    \\
(480, 1[24]) & 2.290 981 375 205 552 301 342 544 968 555 621 408                       & 3.8[26]    \\
(540, 1[27]) & 2.290 981 375 205 552 301 342 544 968 555 621 401 899 8                 & 5.1[29]    \\
(600, 1[30]) & 2.290 981 375 205 552 301 342 544 968 555 621 401 899 680 438           & 6.8[32]    \\
(660, 1[33]) & 2.290 981 375 205 552 301 342 544 968 555 621 401 899 680 431 809       & 9.2[35]    \\
$\infty$     & 2.290 981 375 205 552 301 342 544 968 555 621 401 899 680 431 808(1)    &            \\
ref. \cite{goldmanHighAccuracyAtomic2000} & 2.290 981 375 205 552 301 342 51    \\
ref. \cite{huffSimplifiedCalculation}     & 2.290 981 375 205 552 301 342 544 968 6 \\
\hline
\hline
\end{tabular}
\end{table*}

\makegapedcells
\setcellgapes{4pt}
\begin{table*}[!htbp]
\caption{Expectation values of operators entering Bethe logarithm for $1\;{}^1S$, $2\;{}^1S$ and $2\;{}^3S$ states of helium. Units are a.u.}
\label{operators}
\begin{tabular}{C{3.5cm}L{4.5cm}L{4.5cm}L{4.5cm}}
\hline
\hline
$O$  &\quad\quad  $1\;{}^1S$ &\quad\quad  $2\;{}^1S$ &\quad\quad  $2\;{}^3S$ \\
\hline
$H_0$                               & --2.90372437703411956    & --2.14597404605441732    & --2.17522937823679130   \\
$\vec{p}_1\cdot\vec{p}_2/(r_1r_2)$  &   1.08625399             &   0.0866481596           &   0.007541238           \\
$\vec{p}_1 /(r_1r_2) \vec{p}_2$     &   0.7445215835           &   0.0594759842           &   0.0046430             \\
$1/r_1$                             &   1.6883168007           &   1.1354076861           &   1.1546641529          \\
$1/r_1^2$                           &   6.0174088670           &   4.1469390197           &   4.1704455513          \\
$1/(r_1 r_2)$                       &   2.7086554744           &   0.5618614674           &   5.6072963568          \\
$1/(r_1 r_{12})$                    &   1.9209439219           &   0.3406338458           &   0.3226962217          \\
$p_2^2/r_1$                         &   4.4186846287           &   0.9028098764           &   0.7383180898          \\
$\vec{p}_1\cdot\vec{p}_2/r_1$       &   0.3520859987           &   0.0252661985           &   0.0073765079          \\
$\vec{p}_1 /r_1 \vec{p}_1$          &   5.4636086989           &   4.1504287929           &   4.2216357822          \\
$1/r_1^3$                           & --30.92590072            & --22.78282258            & --23.02253514           \\
$1/(r_1^2r_{12})$                   &   8.0034536162           &   1.3487605330           &   1.16459908097         \\
$p_2^2/r_1^2$                       &   14.111960074           &   2.0642849769           &   0.75191286028         \\
$\vec{p}_1\cdot\vec{p}_2/r_1^2$     &   1.486585057            &   0.120461970            &   0.016221281           \\
$\vec{p}_1 /r_1^2 \vec{p}_1$        &   2.183359822            &   16.45920921            &   16.72047946           \\
$1/(r_1^2r_2)$                      &   9.1720937561           &   1.4720141658           &   1.24270427423         \\
\hline
\hline
\end{tabular}
\end{table*}

\makegapedcells
\setcellgapes{4pt}
\begin{table*}[!htbp]
\caption{Convergence study of Bethe logarithm for the ground state of helium. Values within square brackets represent the exponential part with base 10. Units are a.u.}
\label{resBLHe}
\begin{tabular}{L{1.5cm}L{4cm}L{4cm}L{3.5cm}L{1.5cm}}
\hline
\hline
N&\qquad $\lambda=1[6]$  &\qquad $\lambda=1[8]$  &standard form of BL   &\quad $E_{\rm max}$\\
\hline
650  & 4.370 160 8   & 4.370 160 9      &  4.370 157     & 3.8[12]  \\   
1300 & 4.370 160 28  & 4.370 160 230    &  4.370 158 1   & 6.4[12]  \\   
1950 & 4.370 160 276 & 4.370 160 223 2  &  4.370 158 5   & 1.0[13]  \\   
2600 & 4.370 160 275 & 4.370 160 222 92 &  4.370 158 6   & 1.1[13]  \\   
3250 & 4.370 160 275 & 4.370 160 222 93 &  4.370 158 8   & 1.4[13]  \\    
$\infty$ & 4.370 160 2(1) & 4.370 160 222 9(1) & 4.370 159(2)           \\
ref. \cite{YangApplicationofTheHylleraasBsplineBasisSet} & 4.370 160 22(5)       \\
ref. \cite{drakeBetheLogarithmsPs2000} & 4.370 160 218(3)      \\
ref. \cite{yerokhinTheoreticalEnergiesLowlying2010a} & 4.370 160 222 9(1)    \\
ref. \cite{korobovCalculationNonrelativisticBethe2012} & 4.370 160 223 06(2)   \\
ref. \cite{korobovBetheLogarithmHelium2019} & 4.370 160 223 0703(3) \\
\hline
\hline
\end{tabular}
\end{table*}

\makegapedcells
\setcellgapes{4pt}
\begin{table}[!htbp]
\caption{Convergence study of Bethe logarithm for $2\;{}^1S$ and $2\;{}^3S$ states of helium. Units are a.u.}
\label{resBLHe2}
\begin{tabular}{L{1.2cm}L{3.5cm}L{3.5cm}}
\hline
\hline
N&\qquad $\beta_{\rm He}(2\;{}^1S)$  &\qquad $\beta_{\rm He}(2\;{}^3S)$\\
\hline
650  & 4.366 6          & 4.364 1                 \\
1300 & 4.366 412 8      & 4.364 036 9             \\
1950 & 4.366 412 729    & 4.364 036 823           \\
2600 & 4.366 412 726 48 & 4.364 036 820 6         \\
3250 & 4.366 412 726 40 & 4.364 036 820 51        \\
$\infty$ & 4.366 412 726 3(1) & 4.364 036 820 5(1)\\
ref. \cite{YangApplicationofTheHylleraasBsplineBasisSet} & 4.366 412 71(1)     &                      \\
ref. \cite{drakeBetheLogarithmsPs2000} & 4.366 412 72(7)     & 4.364 036 82(1)      \\
ref. \cite{yerokhinTheoreticalEnergiesLowlying2010a} & 4.366 412 726 2(1)  & 4.364 036 820 41(2)  \\
ref. \cite{korobovBetheLogarithmHelium2019} & 4.366 412 726 417(1)& 4.364 036 820 476(1) \\
\hline
\hline
\end{tabular}
\end{table}

\bibliography{A_Simple_approach_for_precision_calculation_of_Bethe_logarithm}

\end{document}